\documentstyle[12pt]{article}

\begin{document}
\begin{titlepage}

\hfill{UM-P-95/48}

\hfill{RCHEP-95/13}

\vskip 1 cm

\centerline{\Large \bf New and improved quark-lepton symmetric models}

\vskip 1.5 cm

\centerline{\large R. Foot\footnote{foot@physics.unimelb.edu.au} and
R. R. Volkas\footnote{rrv@physics.unimelb.edu.au,
U6409503@hermes.ucs.unimelb.edu.au}}

\vskip 5 mm

\noindent
\centerline{\it Research Centre for High Energy Physics}
\centerline{\it School of Physics}
\centerline{\it University of Melbourne}
\centerline{\it Parkville 3052 Australia}

\vskip 1.5 cm

\centerline{Abstract}

\noindent
We show how the use of a see-saw mechanism based on a $3 \times 3$
neutrino mass matrix texture can considerably simplify Higgs
sectors for quark-lepton symmetric models (and for Standard Model
extensions generally). The main theory we discuss also
incorporates a previously considered scenario whereby the
charged lepton of a particular generation is
necessarily less massive than the corresponding
charge $+2/3$ quark, provided intergenerational mixing is small.

\end{titlepage}

It is interesting to speculate about possible symmetries beyond those
found in the Standard Model (SM). Extended symmetries can both
simplify quantum numbers by relating previously disparate fields and
improve predictive power by relating parameters such as fermion
masses.

In recent years, the related
ideas of leptonic colour SU(3)$_{\ell}$ and
discrete quark-lepton symmetry have been investigated \cite{ql}.
Several models
employing these ideas have been constructed, some phenomenological
studies have been undertaken and some
cosmological issues addressed \cite{qletc}\cite{volkas}.

In this paper we will introduce two new quark-lepton symmetric models
that are simpler than those studied hitherto. The important
development common to both will be the use of a non-standard see-saw
mechanism for neutrinos. This will allow a considerable simplification
of the Higgs boson sector.
We begin by reviewing the standard see-saw mechanism \cite{seesaw}
and then
discussing why this idea can lead to inelegant Higgs boson
sectors in some SM extensions.

The standard see-saw mass matrix for
neutrinos is given by the Lagrangian ${\cal L}_{\nu}$ where
\begin{equation}
{\cal L}_{\nu} = \left[ \begin{array}{cc}
\overline{(\nu_L)^c}\ & \ \overline{\nu}_R
\end{array} \right]
\left( \begin{array}{cc} 0\ & \ m \\ m\ & \ M \end{array} \right)
\left[ \begin{array}{c} \nu_L \\ (\nu_R)^c \end{array} \right]
+ {\rm H.c.}
\end{equation}
The parameter $m$ is the usual neutrino Dirac mass, while $M$ is a
Majorana mass for the right chiral projection. The Majorana mass for
the left-handed neutrino is zero by electroweak invariance for simple
Higgs sectors. The
mass eigenstate fields are two Majorana fermions.
When $M \gg m$, the eigenvalues are approximately given by $m^2/M$ and
$M$ while the eigenstate fields are roughly given by
$\nu_L + (m/M)(\nu_R)^c$ and $(\nu_R)^c - (m/M)\nu_L$ respectively.
The light eigenstate is thus identified with the standard neutrino.
The general philosophy is then to relate the large mass $M$ to a high
symmetry breaking scale thereby providing motivation for the $M \gg m$
limit, and thus also for the suppressed eigenvalue $m^2/M$.

This mechanism can be implemented in any extension of the SM which
employs right-handed neutrinos. A well known example is the left-right
symmetric model \cite{lr}
where the scale $M$ is set by the vacuum expectation
value of the Higgs field that spontaneously breaks the right-sector
weak-isospin gauge group SU(2)$_R$ \cite{triplet}.
Another is the Pati-Salam model \cite{ps}
where $M$ is related to the breakdown scale of the SU(4) colour group
[and usually also SU(2)$_R$] \cite{decuplet}.
In both cases the Higgs multiplets
required are in higher-dimensional representations of SU(2)$_R$ and
SU(4), the triplet and decuplet representations respectively.

The result that the Higgs field $\Delta$ determining $M$ is rather
complicated follows generally from the fact the a Majorana mass term
is derived from $\overline{(\nu_R)^c}\nu_R\Delta$. If $\nu_R$ is in
the fundamental representation, then $\Delta$ will generally have to
be in a higher-dimensional representation. In quark-lepton symmetric
models, $\nu_R$ is located inside an SU(3)$_{\ell}$ triplet while
$\Delta$ is an antisextet \cite{ql}\cite{qletc}.
This state of affairs is rather unfortunate from a model-building
point of view. We would like to keep Higgs sectors simple for reasons
of elegance.

There is an alternative see-saw mechanism that removes the need for
Higgs bosons in high dimensional representations. The idea is to
introduce a sterile fermion $S_L$ which couples to
right-handed neutrinos through a simpler Higgs multiplet $\chi$. The
neutrino mass Lagrangian is then
\begin{equation}
{\cal L}_{\nu} = \left[ \begin{array}{ccc}
\overline{(\nu_L)^c}\ & \ \overline{\nu}_R\ & \ \overline{(S_L)^c}
\end{array} \right]
\left( \begin{array}{ccc} 0\ & \ m\ & 0 \\ m\ & \ 0\ &\ M \\
0\ & \ M\ & \ M_S \end{array} \right)
\left[ \begin{array}{c} \nu_L \\ (\nu_R)^c \\ S_L
\end{array} \right]
+ {\rm H.c.}
\label{3x3}
\end{equation}
where the mass $M$ is now proportional to the VEV of $\chi$. The mass
$m$ is still the electroweak neutrino Dirac mass, while $M_S$ is an
optional bare Majorana mass for $S_L$.

To see the implications of this mass matrix, first set $M_S = 0$.
Diagonalisation of the
mass matrix reveals one massless Weyl neutrino $\nu_{\ell L}$
together with a
Dirac partner $n$ of mass $\sqrt{m^2 + M^2}$. The mass eigenstate fields
are
\begin{eqnarray}
\nu_{\ell L} & = & \cos\alpha \nu_L - \sin\alpha S_L,\nonumber\\
n_L & = & \sin\alpha \nu_L + \cos\alpha S_L,\nonumber\\
n_R & = & \nu_R,
\end{eqnarray}
where the mixing angle $\alpha$ is given by $\tan\alpha = m/M$.
The masslessness of $\nu_{\ell L}$ is related to the presence of an
unbroken lepton number symmetry that is in turn related to the absence
of a bare mass for $S_L$.
If $M_S$ is nonzero but small, then the Weyl state gets transformed
into a Majorana neutrino of approximate mass $m^2 M_S/M^2$.

There are three noteworthy features of this ``$3 \times 3$ see-saw
mechanism.'' First, and of most interest to us here, the Higgs field
$\chi$ has a simple multiplet assignment. Because it couples to
$\overline{\nu}_R S_L$ where $S_L$ has the gauge quantum numbers of
the vacuum, $\chi$ will have the same quantum numbers as the multiplet
of which $\nu_R$ is a member. Second, it can furnish massless
neutrinos if that is what is desired. Third, when massive neutrinos
are desired, it can furnish a doubly-suppressed eigenvalue
$m(m/M)(M_S/M)$ provided that $M_S \ll M$. This means that the high
symmetry breaking scale associated with $M$ can be reduced, leading to
a more easily testable theory.

We now employ this mechanism in the construction of simpler
quark-lepton symmetric models.
Our first decision about the structure of the
model concerns electroweak mass relations. We know from past studies
that discrete quark-lepton symmetry produces the tree-level mass
relations $m_u = m_e$ and $m_d = m^{\rm Dirac}_{\nu}$ provided two
conditions are met. The first condition is that the high scale
symmetry breaking process leave unbroken an SU(2)$'$
subgroup of leptonic colour SU(3)$_{\ell}$ \cite{volkas}.
The second condition is that the electroweak Higgs
boson sector consist of the minimal configuration of a single
electroweak doublet \cite{ql}\cite{qletc}.
Although the $m_d = m^{\rm Dirac}_{\nu}$ relation need
not be a phenomenological concern because of the see-saw mechanism,
the $m_u = m_e$ relation is a problem. In the first model we will
present we choose not to comply with the first condition. In the
second model, we choose not to comply with the second condition.

We now present the first model. To evade the first condition we
arrange for SU(3)$_{\ell}$ to be completely broken, an idea
first discussed in Ref.\cite{volkas}. The model we present achieves
the same end as Ref.\cite{volkas} but in a simpler way.

The gauge group is $G_{q\ell}$ where
\begin{equation}
G_{q\ell} = {\rm SU}(3)_{\ell} \otimes {\rm SU}(3)_q \otimes {\rm
SU}(2)_L \otimes {\rm U}(1)_X.
\end{equation}
The group SU(3)$_q$ is standard quark colour, while the Abelian charge
$X$ is specified by displaying the fermion spectrum of a generation:
\begin{eqnarray}
& Q_L \sim (1,3,2)(1/3),\qquad u_R \sim (1,3,1)(4/3),\qquad d_R
\sim (1,3,1)(-2/3), &\ \nonumber\\
& F_L \sim (3,1,2)(-1/3),\qquad E_R \sim (3,1,1)(-4/3),\qquad N_R \sim
(3,1,1)(2/3), &\ \nonumber\\
& S_L \sim (1,1,1)(0). &\
\end{eqnarray}
The fields $F_L$, $E_R$ and $N_R$ contain the standard left-handed
lepton doublet, right-handed charged lepton and right-handed neutrino,
respectively. The precise location of these fields within the
generalised lepton multiplets will be specified shortly.

The discrete quark-lepton symmetry is given by
\begin{eqnarray}
& Q_L \leftrightarrow F_L,\quad u_R \leftrightarrow E_R,\quad d_R
\leftrightarrow N_R,\quad S_L \leftrightarrow S_L, &\ \nonumber\\
& G_q^{\mu} \leftrightarrow G_{\ell}^{\mu},\quad W_L^{\mu}
\leftrightarrow W_L^{\mu}\quad {\rm and}\quad
C^{\mu} \leftrightarrow -C^{\mu}, &\
\end{eqnarray}
where $G_{q,\ell}^{\mu}$, $W_L^{\mu}$ and $C^{\mu}$ are the gauge boson
multiplets of SU(3)$_{q,\ell}$, SU(2)$_L$ and U(1)$_X$ respectively.

The Yukawa Lagrangian is the sum of an electroweak term ${\cal L}_{\rm
ew}$ and a non-electroweak term ${\cal L}_{\rm non-ew}$ given,
respectively, by
\begin{equation}
{\cal L}_{\rm ew} = \lambda_1 (\overline{Q}_L u_R \phi^c +
\overline{F}_L E_R \phi) + \lambda_2 (\overline{Q}_L d_R \phi +
\overline{F}_L N_R \phi^c) + {\rm H.c.}
\end{equation}
and
\begin{eqnarray}
{\cal L}_{\rm non-ew} & = & h_1 [\overline{(F_L)^c} F_L \chi +
\overline{(Q_L)^c} Q_L \chi'] +  h_2 [\overline{(E_R)^c} N_R \chi +
\overline{(u_R)^c} d_R \chi']\nonumber\\
& + & k_1 [\overline{(F_L)^c} F_L \xi +
\overline{(Q_L)^c} Q_L \xi'] +  k_2 [\overline{(E_R)^c} N_R \xi +
\overline{(u_R)^c} d_R \xi']\nonumber\\
& + & h_3 (\overline{S}_L \chi^{\dagger} N_R + \overline{S}_L
\chi'^{\dagger} d_R) +  k_3 (\overline{S}_L \xi^{\dagger} N_R +
\overline{S}_L \xi'^{\dagger} d_R) + {\rm H.c.}
\end{eqnarray}
The Higgs fields in these equations are given by
\begin{eqnarray}
& \phi \sim (1,1,2)(1),\quad \chi \sim (3,1,1)(2/3),\quad \chi' \sim
(3,1,1)(-2/3), &\ \nonumber\\
& \quad \xi \sim (3,1,1)(2/3)\quad {\rm and}\quad \xi'
\sim (3,1,1)(-2/3). &\
\end{eqnarray}
The transformations
\begin{equation}
\phi \leftrightarrow \phi^c \equiv i\tau_2\phi^*,\qquad
\chi \leftrightarrow \chi'\qquad
{\rm and}\qquad \xi \leftrightarrow \xi'
\end{equation}
define the action of the discrete quark-lepton symmetry on the Higgs
fields.

In the first stage of symmetry breaking, $G_{q\ell}$ breaks to
the SM gauge group $G_{SM}$ where $G_{SM} = $
SU(3)$_q\otimes$SU(2)$_L\otimes$U(1)$_Y$. The weak hypercharge $Y$ is
identified as
\begin{equation}
Y = X + \frac{T_8}{3} - T_3,
\end{equation}
where $T_8$ and $T_3$ are diagonal generators of SU(3)$_{\ell}$ given
by diag$(-2,1,1)$ and diag$(0,1,-1)$ respectively. This symmetry breakdown
pattern is achieved by the most general VEVs for $\chi$ and $\xi$,
namely
\begin{equation}
\langle\chi\rangle = \left( \begin{array}{c} v \\ 0 \\ 0
\end{array} \right)\qquad {\rm and }\qquad
\langle\xi\rangle = \left( \begin{array}{c} w_1 \\ w_2 \\ 0
\end{array} \right).
\end{equation}
It is easy to see that this is the most general pattern: Suppose
$\langle\chi\rangle = (v_1, v_2, v_3)^T$. Then by an SU(2) rotation of
the second and third entries we can set $v_3 = 0$. The VEV is now
$\langle\chi\rangle = (v_1, v_2', 0)^T$. An SU(2) rotation
of the first and second entries then allows us to set $v_2' = 0$,
leaving only the first entry non-zero. We can now follow the same
procedure for $\xi$, but we cannot perform the second SU(2) rotation
since it will in general spoil the $(v, 0, 0)^T$ pattern for
$\langle\chi\rangle$.

Please observe that although $\chi$ and $\xi$ have identical quantum
numbers, the addition of the second leptonically coloured Higgs boson
multiplet drastically changes the qualitative physics of the theory.
If only one of $\chi$ or $\xi$ were to be used, then an unbroken
SU(2)$'$ subgroup of leptonic colour
would necessarily exist. As well as
having implications for electroweak mass relations, this would also
completely change the phenomenology. With completely broken leptonic
colour, all of the lepton-like states have integer charges and are
unconfined. If SU(2)$'$ is unbroken, however, then two of the leptonic
colours become confined charge $\pm 1/2$ fermions. The role of $\chi$
and $\xi$ is thus quite different to the role ascribed to the two
electroweak doublets in a two-Higgs-doublet model. The most general
VEV pattern in the two-Higgs-doublet model would result in
spontaneously broken electromagnetism! One therefore has to choose
parameter space so that this does not happen. By contrast, we will
exploit the greater symmetry breaking capacity of the repeated Higgs
multiplets $\chi$ and $\xi$.

The second stage of symmetry breaking sees the electroweak group
broken by
\begin{equation}
\langle\phi\rangle = \left( \begin{array}{c} 0 \\ u \end{array}
\right)
\end{equation}
where $u \neq 0$. For phenomenological reasons we require that $u \ll
v, w_1, w_2$.

We now need to construct the charged lepton and neutrino mass
matrices. To do this we first need some
notation. Let the leptonic colour components of the generalised leptons
be denoted thus:
\begin{equation}
F_L = \left( \begin{array}{c} \ell_{1L} \\ \ell_{2L} \\ (f_R)^c
\end{array} \right),\qquad
E_R = \left( \begin{array}{c} e_{1R} \\ e_{2R} \\ (\nu_{3L})^c
\end{array} \right)\qquad {\rm and}\qquad
N_R = \left( \begin{array}{c} \nu_{1R} \\ \nu_{2R} \\ (e_{3L})^c
\end{array} \right).
\end{equation}
The weak hypercharges of these components are given by
\begin{equation}
Y(F_L) = \left( \begin{array}{c} -1 \\ -1 \\ +1 \end{array}
\right),\qquad
Y(E_R) = \left( \begin{array}{c} -2 \\ -2 \\ 0 \end{array}
\right)\qquad {\rm and}\qquad
Y(N_R) = \left( \begin{array}{c} 0 \\ 0 \\ +2 \end{array}
\right)
\end{equation}
which justifies the notation. We further denote the weak isospin
components of $\ell_{1L}$, $\ell_{2L}$ and $(f_R)^c$ by
\begin{equation}
\ell_{1L} = \left( \begin{array}{c} \nu_{1L} \\ e_{1L} \end{array}
\right),\qquad
\ell_{2L} = \left( \begin{array}{c} \nu_{2L} \\ e_{2L} \end{array}
\right)\qquad {\rm and}\qquad
(f_R)^c = \left( \begin{array}{c} (e_{3R})^c \\ (\nu_{3R})^c \end{array}
\right),
\end{equation}
where the notation is once again self-evident.

The multiplet pattern above shows that per generation there is one
standard chiral set of leptons together with a lepton--mirror-lepton
pair. We expect that after the first stage of symmetry breaking, the
mirror pair of charged leptons will combine to form a massive Dirac
fermion (as is allowed by the effective $G_{SM}$ symmetry). The
remaining chiral state should have the properties of SM electrons. A
similar phenomenon should occur for neutrinos, but in their case the
see-saw mechanism also occurs as a necessary complication. We will
determine by explicit computation that indeed the lightest charged
lepton and neutrino eigenstates enjoy standard electroweak
interactions (up to small deviations suppressed by the ratio of the
electroweak scale to the quark-lepton symmetry breaking scale).

The charged lepton mass matrix is contained in the Lagrangian ${\cal
L}_e$ where
\begin{equation}
{\cal L}_e = \left( \begin{array}{ccc}
\overline{e}_{1L}\ & \overline{e}_{2L}\ & \overline{e}_{3L}
\end{array}\right)
\left( \begin{array}{ccc}
m_u\ & 0\ & M_3 \\ 0\ & m_u\ & M_1 \\ M_4\ & M_2\ & m_d \end{array}
\right)
\left( \begin{array}{c} e_{1R} \\ e_{2R} \\ e_{3R} \end{array} \right)
+ {\rm H.c.}
\end{equation}
The various entries in this mass matrix are given by,
\begin{eqnarray}
m_u & = & \lambda_1 u,\nonumber\\
m_d & = & \lambda_2 u,\nonumber\\
M_1 & = & 2 h_1 v + 2 k_1 w_1,\nonumber\\
M_2 & = & h_2 v + k_2 w_1,\nonumber\\
M_3 & = & -2 k_1 w_2,\nonumber\\
M_4 & = & -k_2 w_2.
\end{eqnarray}
(For clarity and simplicity we have ignored generation structure
in the above and taken all the Yukawa coupling constants and VEVs
to be real.)
Because of the VEV hierarchy $v, w_1, w_2 \gg u$ we generally expect
that $M_{1,2,3,4} \gg m_{u,d}$. In the limit that the electroweak
masses $m_{u,d} = 0$, the mass matrix produces a pair of massless Weyl
fermions together with two very massive states with masses
$\sqrt{M_1^2 + M_3^2}$ and $\sqrt{M_2^2 + M_4^2}$. The determinant of
the mass matrix is $m_u (M_1 M_2 + M_3 M_4 - m_u m_d) \simeq m_u (M_1
M_2 + M_3 M_4)$. Therefore when the electroweak masses are nonzero,
the massless Weyl pair turn into a Dirac fermion with mass $m_e$ where
\begin{equation}
m_e \simeq m_u \frac{M_1 M_2 + M_3 M_4}{\sqrt{M_2^2 +
M_4^2}\sqrt{M_1^2 + M_3^2}} \equiv m_u \cos(\beta_1 - \beta_2).
\end{equation}
The angles $\beta_{1,2}$ are defined by
\begin{equation}
\tan{\beta_1} \equiv M_2/M_4\qquad {\rm and}\qquad \tan{\beta_2}
\equiv M_1/M_3.
\label{memu}
\end{equation}
We want to identify this light mass eigenstate as the physical
electron. Equation (\ref{memu}) shows that $m_e$ is necessarily less
then $m_u$, given the phenomenologically required VEV hierarchy. This
pleasing result was first considered in a related
quark-lepton symmetric model in Ref.\cite{volkas}. We note that this
qualitative result may be violated if intergenerational mixing is too
large.

To show that our putative electron is a sensible candidate, we must
show that it has approximately the correct electroweak interactions.
Let us return to the $m_{u,d} = 0$ limit. The mass eigenstate fields
are then as follows:
\begin{eqnarray}
& {\rm electron}\ e:\quad
e_L = \sin\beta_2 e_{1L} - \cos\beta_2
e_{2L},\quad e_R = \sin\beta_1 e_{1R} - \cos\beta_1
e_{2R}; &\ \\
& {\rm mass}\ \sqrt{M_1^2 + M_3^2}\ {\rm fermion}\ \epsilon:\quad
\epsilon_L = \cos\beta_2 e_{1L} + \sin\beta_2 e_{2L},\quad
\epsilon_R = e_{3R}; &\ \\
& {\rm mass}\ \sqrt{M_2^2 + M_4^2}\ {\rm fermion}\ \epsilon':\quad
\epsilon'_L = e_{3L},\quad \epsilon'_R = \cos\beta_1
e_{1R} + \sin\beta_1 e_{2R}.
\end{eqnarray}
Note that the left-handed electron is a combination of $e_{1L}$ and
$e_{2L}$ both of which are members of $Y=-1$ SU(2)$_L$ doublets.
Similarly, the right-handed electron is a combination of $e_{1R}$ and
$e_{2R}$ which both have standard electroweak interactions. Our
putative electron will thus have the correct phenomenology.

When the electroweak masses $m_{u,d}$ are switched on, the physical
electron will have small admixtures of $e_{3L}$ and $e_{3R}$ which do
not have standard electroweak assignments. This will result in small
non-standard pieces in the interaction of the physical electron with
$W$ and $Z$ bosons. A phenomenological upper bound on $m/M$ will
result from considerations of weak-interaction nonuniversality and so
on. The derivation of these bounds is beyond the scope of this paper
but will be undertaken in future work \cite{future}.

We now turn to the neutrino sector. It is simplest to first switch off
the mixing of $\nu_{1,2,3}$ with $S_L$. In that case, the neutrino
mass Lagrangian is the same as the charged lepton one with $M_i$
replaced by $-M_i$, $e_i$ replaced by $\nu_i$
and with $m_u$ and $m_d$ interchanged.
There is one light mass eigenstate $\nu$ of approximate mass
$m_d\cos(\beta_1-\beta_2)$ and two heavy eigenstates $n$ and $n'$ with
approximate masses $\sqrt{M_1^2 + M_3^2}$ and $\sqrt{M_2^2 + M_4^2}$,
respectively. The expressions for these fields in the $m_{u,d}=0$
limit are identical to the corresponding expressions for charged
leptons. In particular, $\nu_L$ is primarily composed of $\nu_{1L}$
and $\nu_{2L}$ which both have standard electroweak assignments. The
field $\nu_L$ will thus be a sensible candidate for the physical
left-handed neutrino provided a see-saw mechanism can alter its mass
into a phenomenologically acceptable value.

The effects of the see-saw mechanism are revealed by turning on the
mixing with $S_L$. This mixing is induced by the Yukawa interactions
$\overline{S}_L \chi^{\dagger} N_R$ and $\overline{S}_L \xi^{\dagger}
N_R$. More particularly, these terms mix $S_L$ with $\nu_R$ and
$n'_R$. In the $m_{u,d}=0$ limit, the mass matrix is given through
${\cal L}_{\nu}$ where
\begin{equation}
{\cal L}_{\nu} =
\left[ \begin{array}{cccc}
\overline{n}'_L\ & \overline{S}_L\ & \overline{(n'_R)^c}\ &
\overline{(\nu_R)^c} \end{array} \right]
\left( \begin{array}{cccc}
0\ & 0\ & M_{24} & 0 \\ 0\ & M_S & M_5 & M_6 \\
M_{24} & M_5 & 0\ & 0\ \\ 0\ & M_6 & 0\ & 0
\end{array} \right)
\left[ \begin{array}{c}
(n'_L)^c \\ (S_L)^c \\ n'_R \\ \nu_R \end{array} \right]
\label{4Maj}
\end{equation}
where
\begin{eqnarray}
M_{24} & \equiv & \frac{1}{2}\sqrt{M_2^2 + M_4^2},\nonumber\\
M_5 & \equiv & (h_3 v + k_3 w_1) \cos\beta_1 + k_3 w_2 \sin\beta_1,
\nonumber\\
M_6 & \equiv & (h_3 v + k_3 w_1) \sin\beta_1 - k_3 w_2 \cos\beta_1.
\end{eqnarray}
This produces four very massive Majorana fermions, one of which is the
right-handed neutrino. We conclude,
therefore, that in the $m_{u,d}=0$ limit, the spectrum consists of a
massless Weyl fermion $\nu_L$, a very massive Dirac fermion $n$ and
four very massive Majorana fermions. The massless state $\nu_L$ is
thus an appropriate candidate for the physical light neutrino.

When $m_{u,d}$ are nonzero, the Weyl fermion $\nu_L$ turns into a
Majorana fermion with a small mass. The exact eigenstate field will
now have admixtures other than just $\nu_{1L}$ and $\nu_{2L}$. This
will alter its electroweak properties by a small amount. Once again,
phenomenological bounds constraining this admixture will exist, and
these will be considered in later work. One should in addition note
that the $\nu_L$ mass goes to zero if $M_S=0$ [and the four very
massive Majorana fermions produced by Eq.~(\ref{4Maj}) turn into two
very massive Dirac fermions]. This can be seen either by explicitly
computing the determinant of the full $7 \times 7$ neutrino mass
matrix or by the following qualitative argument: Per generation there
are seven Weyl neutrino-like states. If the Majorana mass
$M_S$ is absent, then all of the mass terms are Dirac-like. The seven
Weyl states can then at most supply three Dirac fermions. The remaining
neutrino field must be Weyl rather than Majorana because there are no
Majorana masses in the mass matrix. Another way to say it is that
setting $M_S$ equal to zero adds an unbroken lepton number
symmetry to the model which forbids a Majorana mass for the
left-over Weyl state. The $7 \times 7$
neutrino mass matrix of our model therefore behaves in a qualitatively
identical fashion to the $3 \times 3$ see-saw mass matrix discussed in
our introductory paragraphs. Our model therefore adheres to the basic
philosophy of this mechanism while being different in detail due to
the extra neutrino degrees of freedom entailed by leptonic colour
SU(3)$_{\ell}$. In particular, it is clear that the double see-saw
suppression of the smallest neutrino mass eigenvalue will also occur
here if $M_S$ is sufficiently small.

This completes our description of the first model.
We now need only sketch the construction of the second
model. This time we
introduce only one leptonic colour triplet Higgs boson
$\chi$ but we have two electroweak Higgs doublets $\phi_1$ and
$\phi_2$. The second model therefore leaves an SU(2)$'$ subgroup of
SU(3)$_{\ell}$ unbroken. The weak hypercharge generator is now $Y = X
+ T_8/3$, and the $T_8 = 1$ fermions that previously formed a
lepton--mirror-lepton pair now become SU(2) doublets of charge $\pm
1/2$ liptons (to introduce nomenclature adopted in previous papers
\cite{qletc}). The
$T_8 = -2$ fermions are the standard leptons. This scenario is
identical to that proposed in most previously considered quark-lepton
symmetric models. We refer the reader to the relevant papers for more
details \cite{2higgs}.

After the first stage of symmetry breaking induced by $\langle\chi\rangle
\neq 0$, the liptons acquire large masses while the leptons, being
chiral under the effective $G_{SM}$, are massless. Leptons become
massive only after the electroweak symmetry breakdown initiated by
nonzero VEVs for $\phi_1$ and $\phi_2$. By making both Higgs doublets
couple to fermion bilinears, $m_u$, $m_d$, $m_e$ and $m^{\rm
Dirac}_{\nu}$ are all free parameters (see Ref.\cite{2higgs}
for details).

The $\overline{S}_L \chi^{\dagger} N_R$ mixing term between $S_L$ and
$\nu_R$ then generates the large mass $M$ in Eq.(\ref{3x3}). The
second model therefore features precisely the $3 \times 3$ see-saw
mechanism discussed at the outset.

We conclude, therefore, by noting that we have successfully simplified
the Higgs sector of quark-lepton symmetric models in a manner which
also produces acceptable quark and lepton masses. The phenomenology of
the first model is similar to that of Ref.\cite{volkas}
in the charged lepton sector,
while being completely different from those
earlier quark-lepton symmetric
models which had an unbroken SU(2)$'$ subgroup of leptonic colour
SU(3)$_{\ell}$ \cite{ql}\cite{qletc}.
The neutrino sector utilises a different but very
interesting see-saw mechanism which distinguishes it both from
Ref.\cite{volkas} and all earlier models.
We will return to phenomenological
bounds derivable from violations of weak universality and the like in
future work \cite{future}.
The second model is essentially a re-interpretation of
the original quark-lepton symmetric model when augmented by a
singlet fermion per generation and a second electroweak Higgs
doublet.

\vspace{2cm}

\centerline{\bf Acknowledgements}

This work was supported by the Australian Research Council. RRV would
like to thank D. S. Shaw for discussions concerning the second model
presented in this paper.

\end{document}